\documentclass[pra,amsmath,amssymb,superscriptaddress]{revtex4}

\usepackage{graphicx}
\usepackage{latexsym}
\usepackage{amsmath,amssymb}
\usepackage{verbatim,times,bbm}

\newcommand{\beq}{\begin{equation}}
\newcommand{\eeq}{\end{equation}}
\newcommand{\beqa}{\begin{eqnarray}}
\newcommand{\eeqa}{\end{eqnarray}}

\begin{document}

\title{Qubit dynamics in a $q$-deformed oscillators environment}

\author{Sonia L'Innocente}
\affiliation{Dipartimento di Matematica \& Informatica,
Universit\`{a} di Camerino, I-62032 Camerino, Italy}
\author{Cosmo Lupo}
\affiliation{Dipartimento di Fisica, Universit\`a di Camerino,
I-62032 Camerino, Italy}
\author{Stefano Mancini}
\affiliation{Dipartimento di Fisica, Universit\`a di Camerino,
I-62032 Camerino, Italy} \affiliation{INFN, Sezione di Perugia,
I-06123 Perugia, Italy}



\begin{abstract}
We study the dynamics of one and two qubits plunged in a
$q$-deformed oscillators environment. Specifically we evaluate the
decay of quantum coherence and entanglement in time when passing
from bosonic to fermionic environments. Slowing down of decoherence
in the fermionic case is found. The effect only manifests at finite
temperature.
\end{abstract}


\maketitle

\section{Introduction}\label{intro}

Open system dynamics is of uppermost importance in the quantum
regime where non classical phenomena turn out to be very fragile
with respect to any noise source. The noise effects are often
modeled as the result of an interaction of the system with a large
number of uncontrollable degrees of freedom, i.e.\ an
\emph{environment} \cite{gardiner}. Environments can be assumed as
to be composed by different kinds of particles, for instance
oscillators or spin-$\frac{1}{2}$. These objects come, under the
mathematical point of view, from the realizations of two different
algebras (the Heisenberg-Weyl algebra and the Lie algebra su$(2)$)
corresponding to fermionic and bosonic commutation relations. These
latter can be seen as two limit cases of more general commutation
relations involving deformed algebras parameterized by one
continuous parameter \cite{Kury, Kulish,Mac}.

Our aim is to analyze the qubit dynamics in an environment of
oscillators satisfying suitable $q$-deformed commutation relations,
such that it permits to continuously interpolate between oscillators
and spin-$\frac{1}{2}$. Actually, we investigate how quantum
decoherence phenomena changes in passing from bosonic to fermionic
environments. We find a slowing down of decoherence in the fermionic
case. However, this effect only manifests at finite temperature.

The paper is organized as follows. In Section \ref{model} we present
the model. We then derive the master equation in Section
\ref{equation}. In Section \ref{onequbit} we study the dynamics of a
single qubit and we evaluate its coherence decay. We then study the
dynamics of two qubits and we evaluate the entanglement decay by
distinguishing the case of the two qubits in the same environment
(Section \ref{twoqubit_same}), from that of the two qubits in
separate environments (Section \ref{twoqubit_separate}). Finally,
Section \ref{conclude} is for concluding remarks.


\section{The model}\label{model}

Let us consider a system (qubit) described by the free Hamiltonian
\begin{eqnarray}
H_S&=&\Omega\sigma_z\label{HS},
\end{eqnarray}
with $\Omega$ the qubit frequency and $\sigma$, $\sigma^{\dag}$, $\sigma_z$ operators
satisfying the commutation relations
\begin{eqnarray}
\left[\sigma^{\dag},\sigma\right]&=&\sigma_z,\\
\left[\sigma,\sigma_z\right]&=&2\sigma,\\
\left[\sigma^{\dag},\sigma_z\right]&=&-2\sigma^{\dag}.
\end{eqnarray}
They define the su$(2)$ algebra. Furthermore, we consider an
environment composed by an infinite (countable) number of
oscillators whose Hamiltonian reads as \cite{Greenberg, Ham1}
\begin{eqnarray}
H_E & = & \sum_k \omega_k N_k, \label{HB1}
\end{eqnarray}
with $\omega_k$ the frequency of the $k$-th oscillator and
$A_k$, $A_k^\dag$, $N_k$ operators satisfying the commutation relations
\begin{eqnarray}
{[} N_h , A_k {]} & = & - \delta_{hk} A_k, \\
{[} N_h , A^\dag_k {]} & = & \delta_{hk} A^\dag_k.
\end{eqnarray}
They define the Heisenberg-Weyl algebra. We are now going to
introduce a deformation of this algebra through the so-called
``quons" commutation relations \cite{Greenberg}
\begin{equation}
A_h A_k^\dag - q A_k^\dag A_h = \delta_{hk},
\end{equation}
where $q \in [-1 , 1]$ is the deformation parameter. It allows us to
interpolate between fermions ($q=-1$) and bosons ($q=1$).
Intermediate values of $q \in (-1,1)$ correspond to the so-called
``infinite statistics".

We assume the system interacting with the environment through the
following Hamiltonian
\begin{eqnarray}
H_I&=&\sum_k\lambda_k\left(A^{\dag}_k\sigma+A_k\sigma^{\dag}\right),\label{HI}
\end{eqnarray}
where $\lambda_k$ denotes the coupling constant of the system with
the $k$-th environment's oscillator.


\section{Master Equation}\label{equation}

Quite generally, the master equation for the system density operator
$\rho$ can be derived by using the Born-Markov approximation
\cite{gardiner}. Hence, it can be formally written as
\begin{equation}
\dot\rho(t)=-\int_0^{\infty}d\tau{\rm Tr}_E\left\{
\left[H_I(t),\left[H_I(t-\tau),\rho(t)\otimes
\rho_E\right]\right]\right\}, \label{ME}
\end{equation}
where $\rho_E$ is the initial environment density operator and ${\rm
Tr}_E$ denotes the trace over environment degrees of freedom.
Furthermore, it is
\begin{equation}
H_I(t)=e^{\iota(H_S+H_E)t}H_Ie^{-\iota(H_S+H_E)t}.
\label{HIt}
\end{equation}

For the choice of the environment Hamiltonian \eqref{HB1}, the
dynamical equations are formally identical to the undeformed case.
The reason is that the interaction Hamiltonian $H_I(t)$ reads as
follows
\begin{equation}
H_I(t)=\sum_k\lambda_k\left(A^{\dag}_k\sigma
e^{-\iota(\omega_k-\Omega)t}+A_k\sigma^{\dag}
 e^{\iota(\omega_k-\Omega)t}\right),
\end{equation}
by virtue of \eqref{HIt}, \eqref{HI}, \eqref{HB1} and \eqref{HS}.
Therefore, from \eqref{ME}, we can write
\begin{equation}\label{ME1}
\dot\rho(t) = - \int_0^{\infty}d\tau{\rm Tr}_E \sum_{k,l}
\lambda_k\lambda_l \mathcal{F}_{kl}(\rho(t)),
\end{equation}
where
\begin{align}
\mathcal{F}_{kl}(\rho(t)) \, = \, & F_k(t) \, F_l(t-\tau) \, \rho(t)\otimes\rho_E - F_k(t) \, \rho(t)\otimes\rho_E \, F_l(t-\tau) \nonumber\\
& - F_k(t-\tau) \, \rho(t)\otimes\rho_E \, F_l(t) +
\rho(t)\otimes\rho_E \, F_k(t-\tau) \, F_l(t)
\end{align}
and
\begin{equation}
F_k(t) = A^{\dag}_k\sigma
e^{-\iota(\omega_k-\Omega)t}+A_k\sigma^{\dag}
e^{\iota(\omega_k-\Omega)t}.
\end{equation}

We now assume an initial thermal state for the environment at
temperature $T$,
\begin{equation}
\rho_E=Z^{-1}e^{-H_E/T},
\end{equation}
where
\begin{equation}
Z={\rm Tr}_E\{e^{-H_E/T}\},
\end{equation}
is the partition function.

In \eqref{ME1} we have nonzero terms of the form
\begin{align}
{\rm Tr}_E\left\{\rho_EA_k^{\dag}(t) A_l(t-\tau)\right\} & =
\frac{1}{Z} {\rm Tr}_E \left\{
\exp\left[-\frac{1}{2T}\sum_j\omega_jN_j\right]
e^{-\iota\omega_kt}A_k^{\dag}A_le^{\iota\omega_l(t-\tau)} \right\} \nonumber\\
&=\delta_{k,l}\frac{1}{Z}\sum_{n_k}[n_k]_q\exp\left[-\frac{\omega_k[n_k]_q}{2T}\right]
e^{-\iota\omega_k\tau}.
\label{AdagA}
\end{align}
Here we have defined
\begin{equation}
[n]_q=\frac{1-q^n}{1-q},
\end{equation}
as the $q$-deformed number.

Then, neglecting principal values terms, we obtain from
\eqref{AdagA}
\begin{align}
\int_0^{\infty} d\tau \sum_{k,l}\lambda_k\lambda_l {\rm Tr}_E\left\{\rho_EA_k^{\dag}(t) A_l(t-\tau)\right\}e^{\iota\Omega\tau}&=
\int_0^{\infty} d\tau \sum_{k}\lambda_k^2 \langle[N(\omega_k)]_q\rangle_{E}\, e^{-\iota(\omega_k-\Omega)\tau}\nonumber\\
&=\sum_{k}\lambda_k^2 \langle[N(\omega_k)]_q\rangle_{E} \delta(\omega_k-\Omega),
\label{notzero1}
\end{align}
where
\begin{equation}
 \langle[N(\omega_k)]_q\rangle_{E}=\frac{1}{Z}\sum_{n_k}[n_k]_q\exp\left[-\frac{\omega_k[n_k]_q}{2T}\right].
\end{equation}
Moving to the continuum of frequencies for the environment
oscillators, we have
\begin{align}
\sum_{k}\lambda_k^2 \langle[N(\omega_k)]_q\rangle_{E} \delta(\omega_k-\Omega)\rightarrow
\int d\omega \Lambda^2(\omega)      \langle[N(\omega)]_q\rangle_{E}
\delta(\omega-\Omega),
\end{align}
where $\Lambda^2(\omega)$ accounts for the coupling spectrum as well
as for the density of states. As usual, we set
$\Lambda^2(\Omega)=\gamma/2$ to be the damping rate. Moreover, we
get the following distribution \cite{Chaichian_JPA_26,Goodison}
\begin{eqnarray}
\langle [ N ]_q \rangle_{E} \equiv \langle [ N(\Omega) ]_q \rangle_{E}& = & \frac{1}{e^{\Omega/T}-q},\\
\langle [ N + 1 ]_q \rangle_{E} \equiv \langle [ N(\Omega) + 1 ]_q \rangle_{E} & = &
\frac{e^{\Omega/T}}{e^{\Omega/T}-q}.
\end{eqnarray}

In summary, from (\ref{notzero1}), we have
\begin{equation}
\int d\tau\sum_{k,l}\lambda_k\lambda_l
{\rm Tr}_E\left\{\rho_E A^{\dag}_k(t) A_l(t-\tau)\right\}e^{-\iota(\omega_k-\Omega)t+\iota(\omega_l-\Omega)(t-\tau)}=\frac{\gamma}{2} \langle [ N ]_q \rangle_{E}.
\end{equation}
Other nonzero terms in \eqref{ME1} are
\begin{equation}
\int d\tau\sum_{k,l}\lambda_k\lambda_l
{\rm Tr}_E\left\{\rho_E A_k(t) A_l^{\dag}(t-\tau)\right\}e^{\iota(\omega_k-\Omega)t-\iota(\omega_l-\Omega)(t-\tau)}=\frac{\gamma}{2} \langle [ N +1]_q \rangle_{E}.
\end{equation}

We finally arrive at the following master equation for the reduced
system (the qubit):
\begin{equation}
\dot\rho(t) = - \frac{\gamma}{2} \langle [ N ]_q \rangle_{E} \left(
\sigma\sigma^{\dag}\rho(t) - 2 \sigma^{\dag}\rho(t)\sigma +
\rho(t)\sigma\sigma^{\dag} \right) \nonumber\\
- \frac{\gamma}{2} \langle [ N + 1 ]_q \rangle_{E} \left(
\sigma^{\dag}\sigma\rho(t) - 2 \sigma\rho(t)\sigma^{\dag} +
\rho(t)\sigma^{\dag}\sigma \right). \label{MEf}
\end{equation}
This equation explicitly shows that the effect of the q-deformation
is to change the rates of emission, which is proportional to
$\langle [ N ]_q \rangle_{E}$, and the rate of absorption,
proportional to $\langle [ N + 1 ]_q \rangle_{E}$ ( see also
\cite{Goodison}). Notice that for $T=0$, there are no effects coming
from the deformation, because for $N=0$ we simply have $\langle [ 0
]_q \rangle_{E}=0$ and $\langle [ 1 ]_q \rangle_{E}=1$; in other
words, the nonlinear effects introduced by the q-deformation cannot
be observed if the environment transitions only concern the vacuum
and the states with single excitation.


\section{One qubit}\label{onequbit}

Let us consider the operators appearing in Eq.(\ref{MEf}) and
represent them in matrix form in the computational basis
$\{|0\rangle,|1\rangle\}$,
\begin{eqnarray}
\sigma=\left(
\begin{array}{cc}
0&0\\
1&0
\end{array}\right),
\end{eqnarray}
and
\begin{eqnarray}
\rho(t)=\left(
\begin{array}{cc}
a(t)&b_1(t)+\iota b_2(t)\\
b_1(t)-\iota b_2(t)&1-a(t)
\end{array}\right),
\end{eqnarray}
where $a(t)$, $b_1(t)$ and $b_2(t)$ are real functions of time to be
determined.

Inserting the above matrices into Eq.(\ref{MEf}) we get the
following set of differential equations
\begin{eqnarray}
\frac{d}{dt}a&=&-2 (A+B)a+2A,
\label{de1}\\
\frac{d}{dt}(b_1+\iota b_2)&=&-(A+B)(b_1+\iota b_2),
\label{de2}
\end{eqnarray}
where for the sake of simplicity we have set
\begin{eqnarray}
A&=&(\gamma/2)\langle [N]_{q}\rangle_E, \label{A}\\
B&=&(\gamma/2)\langle [N+1]_{q}\rangle_E. \label{B}
\end{eqnarray}

The solutions of the differential equations \eqref{de1}, \eqref{de2}
read
\begin{eqnarray}
a(t)&=&\frac{e^{-2(A+B)t}}{A+B}\left[a(0) B+A\left(a(0)+e^{2(A+B)t}-1\right)\right],\\
b_1(t)&=&b_1(0)e^{-(A+B)t},\label{cohere}\\
b_2(t)&=&b_2(0)e^{-(A+B)t}.
\end{eqnarray}

\begin{figure}
\centering
\includegraphics[width=0.5\textwidth]{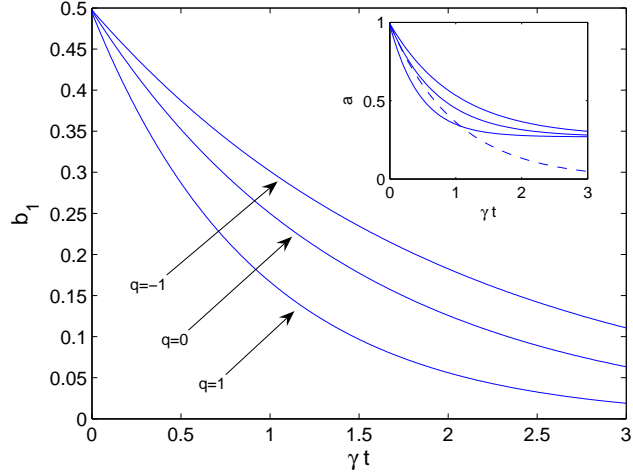}
\caption{The plot shows the decay of the coherence $b_1(t)$ for a
qubit in a quon-environment at temperature $T/\Omega=1$, for
different values of the deformation parameter. In the inset, it is
shown the decay of the population $a(t)$ (solid lines refer again to
$T/\Omega=1$, dashed line refers $T=0$).} \label{coherence}
\end{figure}

Figure \ref{coherence} shows the decay of the coherence ($b_1(t)$)
for a qubit in a quon environment at temperature $T/\Omega=1$, for
different values of the deformation parameter. From Eq.s (\ref{A}),
(\ref{B}), (\ref{cohere}), it follows that the decay of coherence at
$q=-1$, and any finite temperature, behaves as the decay at $T=0$
and any $q$. In the inset, it is shown the decay of the population
$a(t)$ (solid lines refer again to $T/\Omega=1$, while dashed line
refers $T=0$). Thus, the fermionic environment gives rise to the
slowest decay of coherence and population. The decay of quantum
coherence becomes slower and slower when passing from the bosonic to
fermionic environment.

\section{Two qubits in the same environment}\label{twoqubit_same}

We now assume the system composed by two identical qubits
interacting with the same environment. Then the master equation can
be written as Eq.(\ref{MEf}) simply replacing $\sigma$ with
$\sigma_1+\sigma_2$, that is
\begin{eqnarray} \label{MEf2}
\dot\rho(t) & = & -\frac{\gamma}{2} \langle [ N ]_q \rangle_{E} \left( \sigma_1\sigma_1^{\dag}\rho(t) - 2 \sigma_1^{\dag}\rho(t)\sigma_1 + \rho(t)\sigma_1\sigma_1^{\dag} +\sigma_2\sigma_2^{\dag}\rho(t) \right.\nonumber\\
&& \qquad\qquad\quad - 2 \sigma_2^{\dag}\rho(t)\sigma_2 + \rho(t)\sigma_2\sigma_2^{\dag} + \sigma_1\sigma_2^{\dag}\rho(t)+\sigma_2\sigma_1^{\dag}\rho(t) \nonumber\\
&& \qquad\qquad\quad \left. -2\sigma_1^{\dag}\rho(t)\sigma_2-2\sigma_2^{\dag}\rho(t)\sigma_1 +\rho(t)\sigma_1\sigma_2^{\dag}+\rho(t)\sigma_2\sigma_1^{\dag} \right) \nonumber\\
&-& \frac{\gamma}{2} \langle [ N + 1 ]_q \rangle_{E} \left( \sigma_1^{\dag}\sigma_1\rho(t) - 2 \sigma_1\rho(t)\sigma_1^{\dag} + \rho(t)\sigma_1^{\dag}\sigma_1 +\sigma_2^{\dag}\sigma_2\rho(t) \right.\nonumber\\
&& \qquad\qquad\quad - 2 \sigma_2\rho(t)\sigma_2^{\dag} + \rho(t)\sigma_2^{\dag}\sigma_2 + \sigma_1^{\dag}\sigma_2\rho(t) + \sigma_2^{\dag}\sigma_1\rho(t) \nonumber\\
&& \qquad\qquad\quad \left. - 2 \sigma_1\rho(t)\sigma_2^{\dag} - 2
\sigma_2\rho(t)\sigma_1^{\dag} + \rho(t)\sigma_1^{\dag}\sigma_2 +
\rho(t)\sigma_2^{\dag}\sigma_1 \right).
\end{eqnarray}
Then, we proceed in the same way as for the single qubit case. That
is, we consider the operators appearing in Eq.(\ref{MEf2}) and
represent them in matrix form in the computational basis
$\{|00\rangle,|01\rangle|10\rangle,|11\rangle\}$,
\begin{eqnarray}
\sigma_1=\left(
\begin{array}{cccc}
0&0&0&0\\
0&0&0&0\\
1&0&0&0\\
0&1&0&0
\end{array}\right),\quad
\sigma_2=\left(
\begin{array}{cccc}
0&0&0&0\\
1&0&0&0\\
0&0&0&0\\
0&0&1&0
\end{array}\right),
\label{sig12}
\end{eqnarray}
and
\begin{eqnarray}
\rho(t)=\left(
\begin{array}{cccc}
a(t)&b_1(t)+\iota b_2(t)&c_1(t)+\iota c_2(t)&d_1(t)+\iota d_2(t)\\
b_1(t)-\iota b_2(t)&e(t)&f_1(t)+\iota f_2(t)&g_1(t)+\iota g_2(t)\\
c_1(t)-\iota c_2(t)&f_1(t)-\iota f_2(t)&h(t)&i_1(t)+\iota i_2(t)\\
d_1(t)-\iota d_2(t)&g_1(t)-\iota g_2(t)&i_1(t)-\iota i_2(t)&1-a(t)-e(t)-h(t)
\end{array}\right),
\label{rhomat}
\end{eqnarray}
where $a(t)$, $b_1(t)$, $b_2(t)$, $c_1(t)$, $c_2(t)$, $d_1(t)$,
$d_2(t)$, $e(t)$, $f_1(t)$, $f_2(t)$, $g_1(t)$, $g_2(t)$, $h(t)$,
$i_1(t)$ and $i_2(t)$ are real functions of time to be determined.
In terms of these functions, the master equation is written as a set
of coupled differential equations. They are reported together with
their solutions in Appendix \ref{Asame}.


In this case, a relevant quantity to study is the entanglement
between the two qubits. Specifically we consider the qubits
initialized in one of the four Bell states
\begin{eqnarray}
|\phi_\pm\rangle &=&  2^{-1/2}(|00\rangle \pm |11\rangle),
\label{phipm}\\
|\psi_\pm\rangle &=&  2^{-1/2}(|01\rangle \pm |10\rangle),
\label{psipm}
\end{eqnarray}
and then we investigate how entanglement decays.

We use the concurrence as measure of the degree of entanglement
\cite{Woot}
\begin{equation}
C(\rho(t))=\max\left\{0,\lambda_1(t)-\lambda_2(t)-\lambda_3(t)-\lambda_4(t)\right\},
\end{equation}
where $\lambda_i(t)$'s are, in decreasing order, the nonnegative
square roots of the moduli of the eigenvalues of
$\rho(t)\tilde\rho(t)$ with
\begin{equation}
\tilde\rho(t)=\left(\sigma_{1}-\sigma_1^{\dag}\right)\left(\sigma_{2}-\sigma_2^{\dag}\right)
\rho^*(t)
\left(\sigma_{1}-\sigma_1^{\dag}\right)\left(\sigma_{2}-\sigma_2^{\dag}\right),
\end{equation}
and $\rho^*(t)$ denotes the complex conjugate of $\rho(t)$.

\begin{figure}
\centering
\includegraphics[width=0.5\textwidth]{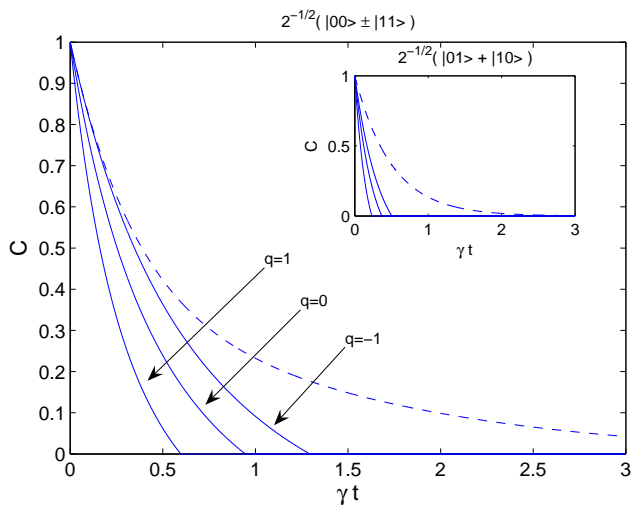}
\caption{The plot shows the decay of the concurrence for the
two-qubit system in a quon-environment. Both qubits are plunged into
the same environment. The qubits are initialized in one of the Bell
states $|\phi_\pm\rangle = 2^{-1/2}(|00\rangle \pm |11\rangle)$;
dashed line refer to $T/\Omega=0$, solid lines to $T/\Omega=1$ and
several values of the deformation parameter $q$. In the inset the
two-qubit state is initialized in the Bell state $|\psi_+\rangle =
2^{-1/2}(|01\rangle + |10\rangle)$. The remaining Bell state
$|\psi_-\rangle = 2^{-1/2}(|01\rangle - |10\rangle)$ is preserved by
the dynamics.} \label{quons_unc1}
\end{figure}

The decay of the concurrence is plotted in Figure \ref{quons_unc1}.
The qubits are initialized in the Bell states \eqref{phipm}. For
$T/\Omega=0$, the decay of the concurrence is independent from the
deformation parameter $q$. For $T/\Omega>0$ we see the phenomenon of
entanglement sudden death \cite{ESD}. We notice however that the
entanglement death time depends on the value of the deformation
parameter $q$. In particular, the slowest decay and the longest
lifetime of entanglement is evident for the fermionic case $q=-1$.
The same happens when the two-qubit state is initialized in
$|\psi_+\rangle = 2^{-1/2}(|01\rangle + |10\rangle)$ (see inset).
The decay of concurrence is slower and slower when continuously
passing from the bosonic to the fermionic environment. On the
contrary, the Bell state $|\psi_-\rangle$ is invariant under the
dynamics of \eqref{MEf2}, thus entanglement in this case is totally
preserved.

\section{Two qubits in separate environments}\label{twoqubit_separate}

Here we consider each of the two identical qubit interacting with
its own environment. Then the master equation is a straightforward
extension of Eq.(\ref{MEf}), that is
\begin{eqnarray} \label{MEf3}
\dot\rho(t) & = & -\frac{\gamma}{2} \langle [ N ]_q \rangle_{E}
\left( \sigma_1\sigma_1^{\dag}\rho(t) - 2
\sigma_1^{\dag}\rho(t)\sigma_1 + \rho(t)\sigma_1\sigma_1^{\dag} +
\sigma_2\sigma_2^{\dag}\rho(t) - 2 \sigma_2^{\dag}\rho(t)\sigma_2 +
\rho(t)\sigma_2\sigma_2^{\dag}
\right) \nonumber\\
&& - \frac{\gamma}{2} \langle [ N + 1 ]_q \rangle_{E} \left(
\sigma_1^{\dag}\sigma_1\rho(t) - 2 \sigma_1\rho(t)\sigma_1^{\dag} +
\rho(t)\sigma_1^{\dag}\sigma_1 +\sigma_2^{\dag}\sigma_2\rho(t) - 2
\sigma_2\rho(t)\sigma_2^{\dag} +
\rho(t)\sigma_2^{\dag}\sigma_2\right).
\end{eqnarray}
It can be solved with the same method of (\ref{MEf2}). The
corresponding differential equations and their solutions are
reported in Appendix \ref{Asepa}.

Figure \ref{quons_unc2} shows the decay of the concurrence in time.
The qubits are initialized in the Bell states \eqref{phipm}. For
$T/\Omega=0$, the concurrence decay is independent from the
deformation parameter $q$. Also in this case, for $T/\Omega>0$, we
see the phenomenon of entanglement sudden death \cite{ESD}. We
notice that the entanglement death time depends on the value of the
deformation parameter $q$. In particular, the slowest decay and the
longest lifetime of entanglement is for the fermionic case $q=-1$.
The same happens when the two-qubit state is initialized in
$|\psi_\pm\rangle = 2^{-1/2}(|01\rangle \pm |10\rangle)$ (see
inset). The decay of entanglement becomes slower and slower when
passing from the bosonic to the fermionic environment. In this case
there is no maximally entangled state that remains invariant under
the dynamics.

\begin{figure}
\centering
\includegraphics[width=0.5\textwidth]{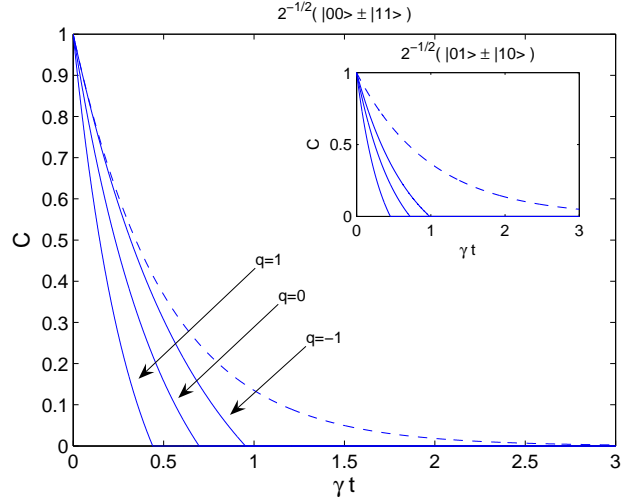}
\caption{The plot shows the decay of the concurrence of the
two-qubit system in a quon-environment. Each qubit is subject to
independent and identical environment. The qubits are initialized in one of
the Bell states $|\phi_\pm\rangle = 2^{-1/2}(|00\rangle
\pm |11\rangle)$; dashed line refer to $T/\Omega=0$, solid lines to
$T/\Omega=1$ and several values of the deformation parameter $q$. In
the inset the two-qubit state is initialized in one of the Bell
states $|\psi_\pm\rangle = 2^{-1/2}(|01\rangle \pm
|10\rangle)$.} \label{quons_unc2}
\end{figure}


\section{Conclusion}\label{conclude}

In conclusion, we have analyzed the qubit dynamics in an environment
of oscillators satisfying suitable $q$-deformed commutation
relations, such that it permits to interpolate between oscillators
and spin-$\frac{1}{2}$ particles. Specifically we have evaluated the
decay of quantum coherence and entanglement in time when passing
from bosonic to fermionic environments. The general behavior is
that, at finite temperature, coherence and entanglement decay slower
and slower when continuously passing from bosonic to fermionic
environments.

Our work sheds further light on the mechanism of loosing quantum
coherence and paves the way for a deeper algebraic analysis of this
phenomenon. Moreover it could be useful for describing realistic
physical situations where the assumption of an interaction with an
environment of solely oscillators (resp. spin-$\frac{1}{2}$)
particle turns out to be oversimplified.

\section*{Acknowledgements}

The work of C.L. and S.M. is partially supported by EU through the
FET-Open Project HIP (FP7-ICT-221899).

\appendix

\section{Two qubits in the same environment}\label{Asame}

Using the parametrization in \eqref{rhomat}, the master equation
\eqref{MEf2} translates in the following set of differential
equations
\begin{eqnarray}
\frac{d}{dt}a & = & -4Ba+2A(e+2f_1+h) \label{de21} \\
\frac{d}{dt}(b_1+\iota b_2) & = & -B(3b_1+3\iota b_2+c_1+\iota c_2) \nonumber\\
&& -A(b_1+\iota b_2+c_1+\iota c_2 -2g_1-2\iota g_2-2i_1-2\iota i_2) \\
\frac{d}{dt}(c_1+\iota c_2) & = & -B(b_1+\iota b_2+3c_1+3\iota c_2) \nonumber\\
&& -A(b_1+\iota b_2+c_1+\iota c_2 -2g_1-2\iota g_2-2i_1-2\iota i_2) \\
\frac{d}{dt}(d_1+\iota d_2) & = & -2(A+B)(d_1+\iota d_2) \\
\frac{d}{dt}e & = & 2B(a-e-f_1)+A(a+2e+h+f_1-1) \\
\frac{d}{dt}(f_1+\iota f_2) & = & B(2a-e-h-2f_1-2\iota f_2)\nonumber\\
&& +A(2-2a-3e-3h-2f_1-2\iota f_2) \\
\frac{d}{dt}(g_1+\iota g_2) & = & B(2b_1+2\iota b_2+2c_1+2\iota c_2-g_1-\iota g_2-i_1-\iota i_2) \nonumber\\
&& -A(3g_1+3\iota g_2+i_1+\iota i_2) \\
\frac{d}{dt}h & = & 2B(a-h-f_1)-2A(a+e+2h+f_1-1) \\
\frac{d}{dt}(i_1+\iota i_2) & = & B(2b_1+2\iota b_2+2c_1+2\iota c_2-g_1-\iota g_2-i_1-\iota i_2) \nonumber\\
&& -A(g_1+\iota g_2+3i_1+3\iota i_2) \label{de29}
\end{eqnarray}
The solutions of the differential equations
\eqref{de21}-\eqref{de29} read


\begin{eqnarray}
a(t) & = & \frac{1}{4\sqrt{B}}e^{-4(\sqrt{AB}+B)t}\left[ 2 a(0)\sqrt{B}\left(1+e^{8\sqrt{AB}t}\right)\right. \nonumber\\
&& \left. +
\sqrt{A}\left(e^{8\sqrt{AB}t}-1\right)\left(e(0)+2f_1(0)+h(0)\right)
\right],
\end{eqnarray}


\begin{eqnarray}
b_1(t) & = & \frac{e^{-\left(3 A+3 B+\Theta\right) t} }{4 \Theta^2}
\left\{A^2 \left[c_1(0) \left(1+e^{2 \Theta t}-2 e^{\left(3
A+B+\Theta \right) t}\right)\right.\right.\nonumber\\
&& \left. + b_1(0) \left(1+e^{2 \Theta t}+2 e^{\left(3
A+B+\Theta\right) t}\right)\right] + B \Theta (b_1(0)+c_1(0)) \left(1-e^{2 \Theta t}\right) \nonumber\\
&& + B^2 \left[ + c_1(0) \left(1+e^{2 \Theta t}-2 e^{\left(3
A+B+\Theta\right) t}\right) + b_1(0) \left(1+e^{2 \Theta t}+2
e^{\left(3 A+B+\Theta\right) t}\right)\right]\nonumber\\
&&\left. +A \left[14 B \left(c_1(0) \left(1+e^{2 \Theta t}-2
e^{\left(3 A+B+\Theta\right) t}\right)+b_1(0) \left(1+e^{2 \Theta
t}+2 e^{\left(3 A+B+\Theta \right) t}\right)\right)
\right.\right.\nonumber\\
&&\left.\left. +\Theta \left(-1+e^{2 \Theta t}\right)
(b_1(0)+c_1(0)+4 g_1(0)+4 i_1(0) )\right] \right\},
\end{eqnarray}


\begin{eqnarray}
c_1(t) & = & \frac{e^{-\left(3 A+3 B+\Theta\right) t} }{4 \Theta^2}
\left\{A^2 \left[b_1(0) \left(1+e^{2 \Theta t}-2 e^{\left(3
A+B+\Theta \right) t}\right)\right.\right.\nonumber\\
&& \left. +c_1(0) \left(1+e^{2 \Theta t}+2 e^{\left(3
A+B+\Theta\right) t}\right)\right] + B \Theta (b_1(0)+c_1(0))
\left(1-e^{2 \Theta t}\right) \nonumber\\
&& +B^2 \left[b_1(0) \left(1+e^{2 \Theta t}-2 e^{\left(3
A+B+\Theta\right) t}\right)+c_1(0) \left(1+e^{2 \Theta t}+2
e^{\left(3 A+B+\Theta\right) t}\right)\right]\nonumber\\
&& +A \left[14 B \left(b_1(0) \left(1+e^{2 \Theta t}-2 e^{\left(3
A+B+\Theta\right) t}\right)+c_1(0) \left(1+e^{2 \Theta t}+2
e^{\left(3 A+B+\Theta \right) t}\right)\right)
\right.\nonumber\\
&&\left.\left. + \Theta \left(-1+e^{2 \Theta t}\right)
(b_1(0)+c_1(0)+4 g_1(0)+4 i_1(0) )\right] \right\},
\end{eqnarray}


\begin{eqnarray}
d_1(t)&=&d_1(0)e^{-2(A+B)t},
\end{eqnarray}


\begin{eqnarray}
e(t) & = & \frac{ e^{-2 \left(3 A+2 \sqrt{AB}+3 B\right) t} }{ 8
\sqrt{AB}\left(A^2-3 A B+B^2\right)} \left\{-2 a(0) AB^2 e^{2 (3
A+B) t} \left(e^{8 \sqrt{AB} t}-1\right)\right.\nonumber\\
&& +2 a(0) B^3 e^{2 (3 A+B) t} \left(e^{8 \sqrt{AB} t}-1\right)\nonumber\\
&& - A^3 e^{2 (3 A+B) t} \left(e^{8 \sqrt{AB} t}-1\right) (e(0)+2 f_1(0)+h(0)) \nonumber\\
&& +3 A^2 B e^{2 (3 A+B) t} \left(e^{8 \sqrt{AB} t}-1\right) (e(0)+2 f_1(0)+h(0))\nonumber\\
&& + \sqrt{A B^5} e^{2 (A+B) t} \left[ 4 e^{4
\left(A+\sqrt{AB}+B\right) t}+4 e^{2
\left(\sqrt{A}+\sqrt{B}\right)^2 t} (e(0)-h(0))\right. \nonumber\\
&& + 2 e^{4 \sqrt{AB} t+4 B t} (e(0)-2 f_1(0)+h(0)-2) \nonumber\\
&& \left. + e^{4 A t} (e(0)+2 f_1(0)+h(0)) + e^{4 A t+8 \sqrt{AB} t} (e(0)+2 f_1(0)+h(0))\right] \nonumber\\
&& -\sqrt{A^3B^3} e^{2 (A+B) t} \left[ 12 e^{4
\left(A+\sqrt{AB}+B\right) t} +12 e^{2
\left(\sqrt{A}+\sqrt{B}\right)^2 t} (e(0)-h(0)) \right. \nonumber\\
&& \left. - 6 a(0) \left(e^{4 A t}+e^{4 A t+8 \sqrt{AB} t}-2 e^{4
\sqrt{AB} t+4 B t}\right) + e^{4 A t} (e(0)+2 f_1(0)+h(0))
\right.\nonumber\\
&& +e^{4 A t+8 \sqrt{AB} t} (e(0)+2 f_1(0)+h(0)) \nonumber\\
&& \left.+2 e^{4 \sqrt{AB} t+4 B t} (5 e(0)-2 f_1(0)+5h(0)-6)\right] \nonumber\\
&&\left. +2 \sqrt{A^5B} \left[-a(0) \left(e^{2 (3 A+B) t}+e^{2
\left(3 A+4 \sqrt{AB}+B\right) t}-2 e^{2 \left(A+2 \sqrt{AB}+3
B\right) t}\right)
\right.\right.\nonumber\\
&& + 2 e^{2 \left(A+2 \sqrt{AB}+2 B\right) t} \left(e^{2 (2 A+B) t}+e^{2 A t} (e(0)-h(0))\right.\nonumber\\
&& \left.\left.\left.+e^{2 B t} (e(0)+h(0)-1)\right)\right]\right\},
\end{eqnarray}


\begin{eqnarray}
f_1(t) & = & \frac{ e^{-4 \left(A+\sqrt{AB}+B\right) t} }{ 8
\sqrt{AB}\left(A^2-3 A B+B^2\right)} \left\{-10 a(0) AB^2 e^{4A t}
\left(e^{8 \sqrt{AB} t}-1\right)\right.\nonumber\\
&& +2 a(0) B^3 e^{4 A t} \left(e^{8 \sqrt{AB} t}-1\right) +A^3 e^{4
A t} \left(e^{8 \sqrt{AB} t}-1\right) (e(0)+2 f_1(0)+h(0)) \nonumber\\
&& +A^2 B e^{4At} \left(e^{8 \sqrt{AB} t}-1\right) (4a(0)-3e(0)-6 f_1(0)-3h(0))\nonumber\\
&& -2 \sqrt{A^5B} \left[2 e^{4 \left(A+\sqrt{AB}+B\right) t}-a(0)
\left(e^{4 A t}-2 e^{4 \left(\sqrt{AB}+B\right) t}+e^{4 A t+8
\sqrt{AB} t}\right)\right.\nonumber\\
&& +2 e^{4 \left(\sqrt{AB}+B\right) t} (e(0)+h(0)-1) \nonumber\\
&&\left.\left.-e^{4 A t} (e(0)+2 f_1(0)+h(0))-e^{4 A t+8 \sqrt{AB}
t} (e(0)+2 f_1(0)+h(0))\right]
\right.\nonumber\\
&& + \sqrt{AB^5} \left[-4 e^{4 \left(A+\sqrt{AB}+B\right) t}-2 e^{4 \left(\sqrt{AB}+B\right) t} (e(0)-2f_1(0)+h(0)-2)\right.\nonumber\\
&& \left. +e^{4 A t} (e(0)+2 f_1(0)+h(0)) + e^{4 A t+8 \sqrt{AB} t} (e(0)+2 f_1(0)+h(0))\right]\nonumber\\
&& + \sqrt{A^3 B^3} \left[12 e^{4 \left(A+\sqrt{AB}+B\right) t}-6
a(0) \left(e^{4 A t}-2 e^{4 \left(\sqrt{AB}+B\right) t}+e^{4 A t+8
\sqrt{AB} t}\right) \right. \nonumber\\
&& -5 e^{4 A t} (e(0)+2 f_1(0)+h(0)) -5 e^{4 A t+8 \sqrt{AB} t} (e(0)+2 f_1(0)+h(0))\nonumber\\
&&\left.\left.+2 e^{4 \left(\sqrt{AB}+B\right) t} (5 e(0)-2f_1(0)+5
h(0)-6)\right]\right\},
\end{eqnarray}


\begin{eqnarray}
f_2(t)&=&f_2(0)e^{-2(A+B)t},
\end{eqnarray}


\begin{eqnarray}
g_1(t) & = & \frac{e^{-\left(3 A+3 B+\Theta\right) t} }{4 \Theta^2}
\left\{B \left[4 \Theta \left(b_1(0)+c_1(0)\right) \left(-1+e^{2
\Theta t}\right)+14 A \left(g_1(0)+i_1(0)\right)\right.\right.\nonumber\\
&& -\Theta \left(g_1(0)+i_1(0)\right) +\Theta e^{2 \Theta t}
\left(g_1(0)+i_1(0)\right)\nonumber\\
&&\left. +28 A e^{\left(A+3 B+\Theta\right) t}
\left(g_1(0)-i_1(0)\right)+14 A e^{2 \Theta t}
\left(g_1(0)+i_1(0)\right)\right]
\nonumber\\
&& +B^2 \left[\left(1+e^{2 \Theta t}+2 e^{\left(A+3 B+\Theta\right)
t}\right) g_1(0)+\left(1+e^{2 \Theta t}-2 e^{\left(A+3
B+\Theta\right) t}\right) i_1(0)\right] \nonumber\\
&& + A \left[\Theta \left(1-e^{2 \Theta t}\right) (g_1(0)+i_1(0))+A
\left(\left(1+e^{2 \Theta t}+2 e^{\left(A+3 B+\Theta\right)
t}\right) g_1(0) \right.\right. \nonumber\\
&& + \left.\left.\left.\left(1+e^{2 \Theta t}-2 e^{\left(A+3
B+\Theta\right) t}\right) i_1(0)\right)\right]
\right\},\nonumber\\
\end{eqnarray}


\begin{eqnarray}
h(t) & = & \frac{ e^{-2 \left(3 A+2 \sqrt{AB}+3 B\right) t} }{ 8
\sqrt{AB}\left(A^2-3 A B+B^2\right)} \left\{- 2 a(0) AB^2 e^{2 (3
A+B) t} \left(e^{8 \sqrt{AB} t}-1\right)\right.\nonumber\\
&& + 2 a(0) B^3 e^{2 (3 A+B) t} \left(e^{8 \sqrt{AB} t}-1\right) \nonumber\\
&& -A^3 e^{2 (3 A+B) t} \left(e^{8 \sqrt{AB} t}-1\right) (e(0)+2 f_1(0)+h(0))\nonumber\\
&& +3 A^2 B e^{2 (3 A+B) t} \left(e^{8 \sqrt{AB} t}-1\right) (e(0)+2 f_1(0)+h(0)) \nonumber\\
&& +\sqrt{A B^5} e^{2 (A+B) t} \left[4 e^{4
\left(A+\sqrt{AB}+B\right) t}-4 e^{2
\left(\sqrt{A}+\sqrt{B}\right)^2 t} (e(0)-h(0))\right.\nonumber\\
&& + 2 e^{4 \sqrt{AB} t+4 B t} (e(0)-2 f_1(0)+h(0)-2) \nonumber\\
&&\left.\left. +e^{4 A t} (e(0)+2 f_1(0)+h(0))+e^{4 A t+8 \sqrt{AB} t} (e(0)+2 f_1(0)+h(0))\right] \right. \nonumber\\
&& - \sqrt{A^3B^3} e^{2 (A+B) t} \left[ 12 e^{4 \left(A+\sqrt{AB}+B\right) t} \right. \nonumber\\
&& - 6 a(0) \left(e^{4 A t}+e^{4 A t+8 \sqrt{AB} t}-2 e^{4 \sqrt{AB} t+4 B t}\right) \nonumber\\
&&-12 e^{2 \left(\sqrt{A}+\sqrt{B}\right)^2 t} (e(0)-h(0)) +e^{4 A t} (e(0)+2 f_1(0)+h(0)) \nonumber\\
&& + e^{4 A t+8 \sqrt{AB} t} (e(0)+2 f_1(0)+h(0)) \nonumber\\
&& \left. +2 e^{4 \sqrt{AB} t+4 B t} (5 e(0)-2 f_1(0)+5 h(0)-6)\right] \nonumber\\
&& \left. +2 \sqrt{A^5B} \left[-a(0) \left(e^{2 (3 A+B) t}+e^{2
\left(3 A+4 \sqrt{AB}+B\right) t}-2 e^{2 \left(A+2 \sqrt{AB}+3 B\right) t}\right) \right.\right.\nonumber\\
&& + 2 e^{2 \left(A+2 \sqrt{AB}+2 B\right) t} \left(e^{2 (2 A+B) t}-e^{2 A t} (e(0)-h(0)) \right. \nonumber\\
&& \left.\left.\left. +e^{2 B t}
(e(0)+h(0)-1)\right)\right]\right\},
\end{eqnarray}


\begin{eqnarray}
i_1(t) & = & \frac{e^{-\left(3 A+3 B+\Theta\right) t} }{4 \Theta^2}
\left\{B \left[4 \Theta \left(b_1(0)+c_1(0)\right) \left(-1+e^{2
\Theta t}\right)\right.\right. \nonumber\\
&& + 14 A \left(g_1(0)+i_1(0)\right)-\Theta \left(g_1(0)+i_1(0)\right) + \Theta e^{2 \Theta t} \left(g_1(0)+i_1(0)\right) \nonumber\\
&&\left.\left. - 28 A e^{\left(A+3 B+\Theta\right) t}
\left(g_1(0)-i_1(0)\right)+14 A e^{2 \Theta t}
\left(g_1(0)+i_1(0)\right)\right] \right.\nonumber\\
&&\left. +B^2 \left[\left(1+e^{2 \Theta t}-2 e^{\left(A+3
B+\Theta\right) t}\right) g_1(0)+\left(1+e^{2 \Theta t}+2
e^{\left(A+3 B+\Theta\right) t}\right) i_1(0)\right]
\right.\nonumber\\
&& + A^2 \left[\left(1+e^{2 \Theta t}-2 e^{\left(A+3 B+\Theta\right)
t}\right) g_1(0)+\left(1+e^{2 \Theta t}+2 e^{\left(A+3
B+\Theta\right) t}\right) i_1(0)\right]  \nonumber\\
&& \left. + A \left[\Theta \left(1-e^{2 \Theta t}\right)
(g_1(0)+i_1(0))\right] \right\},
\end{eqnarray}
where $\Theta=\sqrt{A^2+14AB+B^2}$. The other solutions $b_2(t)$,
$c_2(t)$, $d_2(t)$, $g_2(t)$ and $i_2(t)$ can be obtained from
$b_1(t)$, $c_1(t)$, $d_1(t)$, $g_1(t)$ and $i_1(t)$ respectively by
simply replacing the subscripts $1\to 2$.


\section{Two qubits in separate environments}\label{Asepa}

Using the parametrization in \eqref{rhomat}, the master equation
\eqref{MEf3} translates in the following set of differential
equations
\begin{eqnarray}
\frac{d}{dt}a&=&-4Ba+2A(e+h),
\label{de31}\\
\frac{d}{dt}(b_1+\iota b_2)&=&-3B(b_1+\iota b_2)-A(b_1+\iota b_2-2i_1-2\iota i_2),\\
\frac{d}{dt}(c_1+\iota c_2)&=&-3B(c_1+\iota c_2)-A(c_1+\iota c_2-2g_1-2\iota g_2),\\
\frac{d}{dt}(d_1+\iota d_2)&=&-2(A+B)(d_1+\iota d_2),\\
\frac{d}{dt}e&=&2B(a-e)-2A(a+2e+h-1),\\
\frac{d}{dt}(f_1+\iota f_2)&=&-2(A+B)(f_1+\iota f_2),\\
\frac{d}{dt}(g_1+\iota g_2)&=&B(2c_1+2\iota c_2-g_1-\iota g_2)-3A(g_1+\iota g_2),\\
\frac{d}{dt}h&=&2B(a-h)-2A(a+e+2h-1),\\
\frac{d}{dt}(i_1+\iota i_2)&=&B(2b_1+2\iota b_2-i_1-\iota
i_2)-3A(i_1+\iota i_2). \label{de39}
\end{eqnarray}

The solutions of the differential equations
\eqref{de31}-\eqref{de39} read
\begin{eqnarray}
a(t) & = & \frac{e^{-4(A+B)t}}{(A+B)^2}\left\{ a(0)B^2 +
A^2 \left[a(0)\left(2e^{2(A+B)t}-1\right)\right.\right.\nonumber\\
&& \qquad\qquad\quad\left. + \left(e^{2(A+B)t}-1\right)\left(e^{2(A+B)t}+e(0)+h(0)-1\right)\right] \nonumber\\
&& \qquad\qquad\quad \left. +
\left[2a(0)e^{2(A+B)t}+\left(e(0)+h(0)\right)\left(e^{2(A+B)t}-1\right)
\right]AB\right\},\\
b_1(t)&=&\frac{e^{-3(A+B)t}}{A+B}\left\{b_1(0)B+A\left[b_1(0)e^{2(A+B)t}+
\left(e^{2(A+B)t}-1\right)i_1(0)\right]\right\},\\
c_1(t)&=&\frac{e^{-3(A+B)t}}{A+B}\left\{c_1(0)B+A\left[c_1(0)e^{2(A+B)t}+
\left(e^{2(A+B)t}-1\right)g_1(0)\right]\right\},\\
d_1(t)&=&d_1(0)e^{-2(A+B)t},\\
e(t)&=&\frac{e^{-4(A+B)t}}{(A+B)^2}\left\{
\left[a(0)\left(1-e^{2(A+B)t}\right)+e(0)
-e^{2(A+B)t}\left(h(0)-1\right)+h(0)-1\right]A^2\right.\nonumber\\
&&\left.\qquad\qquad\quad+\left[e(0)+e^{4(A+B)t}+e^{2(A+B)t}\left(e(0)-h(0)-1\right)
+h(0)\right]AB\right.\nonumber\\
&&\left.\qquad\qquad\quad+\left[a(0)\left(e^{2(A+B)t}-1\right)+e(0)e^{2(A+B)t}\right]B^2\right\},\\
f_1(t)&=&f_1(0)e^{-2(A+B)t},\\
g_1(t)&=&\frac{e^{-3(A+B)t}}{A+B}\left\{g_1(0)B+A\left[g_1(0)e^{2(A+B)t}+
\left(e^{2(A+B)t}-1\right)c_1(0)\right]\right\},\\
h(t)&=&\frac{e^{-4(A+B)t}}{(A+B)^2}\left\{
\left[a(0)\left(1-e^{2(A+B)t}\right)-e^{2(A+B)t}\left(e(0)-1\right)+e(0)+h(0)-1\right]A^2\right.\nonumber\\
&&\left.\qquad\qquad\quad+\left[e(0)+e^{4(A+B)t}+h(0)+e^{2(A+B)t}\left(h(0)-e(0)-1\right)\right]AB\right.\nonumber\\
&&\left.\qquad\qquad\quad+\left[a(0)\left(e^{2(A+B)t}-1\right)+h(0)e^{2(A+B)t}\right]B^2\right\},\\
i_1(t)&=&\frac{e^{-3(A+B)t}}{A+B}\left\{i_1(0)B+A\left[i_1(0)e^{2(A+B)t}+
\left(e^{2(A+B)t}-1\right)b_1(0)\right]\right\}.\\
\end{eqnarray}
The other solutions $b_2(t)$, $c_2(t)$, $d_2(t)$, $f_2(t)$, $g_2(t)$
and $i_2(t)$ can be obtained from $b_1(t)$, $c_1(t)$, $d_1(t)$,
$f_1(t)$, $g_1(t)$ and $i_1(t)$ respectively by simply replacing the
subscripts $1\to 2$.


\end{document}